\documentclass[journal]{IEEEtran}

\usepackage[pdftex]{graphicx}
\begin{document}

\title{Light on Dark Matter \\ with Weak Gravitational Lensing}
\author{S.~Pires, J.-L. Starck and A. R\'efr\'egier\\~\IEEEmembership{Laboratoire AIM, CEA/DSM-CNRS-Universite Paris Diderot, IRFU/SEDI-SAP, CEA Saclay,\\Orme des Merisiers, 91191 Gif-sur-Yvette, France}}
\maketitle
\markboth{Light on Dark Matter with Weak Gravitational Lensing}
\maketitle

\begin{abstract}

%Modern cosmology is based on a cosmological model that deals with the structure of the Universe, its origins and its evolution. Over the last decades, observations have provide evidence for both dark matter and dark energy. Universe has been found to be dominated by these two components whose composition remains a mystery. This is not well described or even understood by modern science, so this study could provide the next breakthrough in physics. 

%Modern cosmology is based on a cosmological model that deals with the structure of the Universe, its origins and its evolution. This cosmological model have various parameters that need to be estimated. 

%This paper is a tutorial about the signal processing activity in the weak lensing research area. The field of weak lensing is motivated by the
% that tries to shed light on the dark matter component of the Universe.Over 
This paper reviews statistical methods recently developed to reconstruct and analyze dark matter mass maps from weak lensing observations. The field of weak lensing is motivated by the observations made in the last decades showing that the visible matter represents only about 4-5\% of the Universe, the rest being dark. The Universe is now thought to be mostly composed by an invisible, pressureless matter -potentially relic from higher energy theories- called ``dark matter" (20-21\%) and by an even more mysterious term, described in Einstein equations as a vacuum energy density, called ``dark energy" (70\%). This ``dark" Universe is not well described or even understood, so this point could be the next breakthrough in cosmology.

Weak gravitational lensing is believed to be the most promising tool to understand the nature of dark matter 
%dark energy and then
and to constrain the cosmological model used to describe the Universe. Gravitational lensing is the process in which light from distant galaxies is bent by the gravity of intervening mass in the Universe as it travels towards us. This bending causes the image of background galaxies to appear slightly distorted and can be used to extract significant results for cosmology.

Future weak lensing surveys are already planned in order to cover a large fraction of the sky with large accuracy. However this increased accuracy also places greater demands on the methods used to extract the available information. In this paper, we will first describe the important steps of the weak lensing processing to reconstruct the dark matter distribution from shear estimation. Then we will discuss the problem of statistical estimation in order to set constraints on the cosmological model. We review the methods which are currently used especially new methods based on sparsity.

\end{abstract}

\begin{keywords}
Cosmology : Weak Lensing, Methods : Statistics, Data Analysis
\end{keywords}

%\IEEEpeerreviewmaketitle
\section{Introduction}
According to present observations, we believe that the majority of the Universe is dark, i.e. does not emit electromagnetic radiations.
%cannot be detected from the light which it emits. 
Its presence is inferred indirectly from its gravitational effects: on the motions of astronomical objects and on light propagation.

Weak gravitational lensing has been found to be one of the most promising tools to probe dark matter and dark energy because it provides a method to map directly the distribution of dark matter in the Universe (see \cite{Bartelmann 1999, Mellier 1999}). From this dark matter distribution, the nature of dark matter can be better understood and better constraints can be set on dark energy because it affects the evolution of structures.
This method is now widely used but, the amplitude of the weak lensing signal is so weak that its detection relies on the accuracy of the techniques used  to analyze the data. %Thus, it is necessary to measure the distortion to extremely high accuracy for millions of galaxies, in the presence of observational problems such as blurring, pixelisation and noise. \\
%In this paper, we will present the general problem of weak lensing and we will detail the following sub-problems required for weak lensing data analysis:
%\begin{enumerate}
%\item The measurement of the background galaxy ellipticities to derive a shear map. 
%\item The inversion methods to derive a dark matter mass map from the shear map.
%\item The filtering of the dark matter mass map to reduce the noise level.
%\item The statistical analysis of the weak lensing data to constrain the cosmological model. \\
%\end{enumerate}
Each step of the analysis has required the development of advanced techniques dedicated to these applications.
%The aim of this paper is 
%to present an overview of the currently used methods especially recent techniques based on sparsity.\\

This paper is organized as follow:

Section 2 aims at giving an overview of weak gravitational lensing : the basics of the lensing theory and a brief description of the weak lensing data analysis. 

Section 3 will be dedicated to the presentation of the shear estimation problem. It requires the measurement of the shape of millions of galaxies with extremely high accuracy, in the presence of observational problems such as anisotropic Point Spread Function, pixelisation and noise. Methods that are currently used to derive the lensing shear field from the shapes of background galaxies will be described.

Section 4 will address the following inverse problem : how to derive a dark matter mass map from the measured shear field. Because of some observational effects such as noise and complex survey geometry, this problem can be seen as an ill-posed inverse problem. We will present various methods of inversion currently used to reconstruct the dark matter mass map from incomplete shear map especially a recent promising method of interpolation, based on the sparse representation of the data. And we will describe the different filtering methods which are used to reduce the noise in these dark matter mass maps such as linear filters, Bayesian techniques and a recent wavelet method. 

Finally, in section 5, we will discuss the problem of statistical information extraction in weak lensing data in order to constrain the cosmological model.
We will introduce different statistics which are of interest in weak lensing data analysis. A recent approach based on sparse representations will be presented. \\
%The problem of statistical estimation in incomplete data will also be addressed.\\

%\section{Conclusion}
%The conclusion goes here.

%\appendices

\section{Introduction to weak lensing}
\subsection{Gravitational Lensing observations}
\begin{figure}[h!]
\centerline{
\includegraphics[width=7.cm, height=4.3cm]{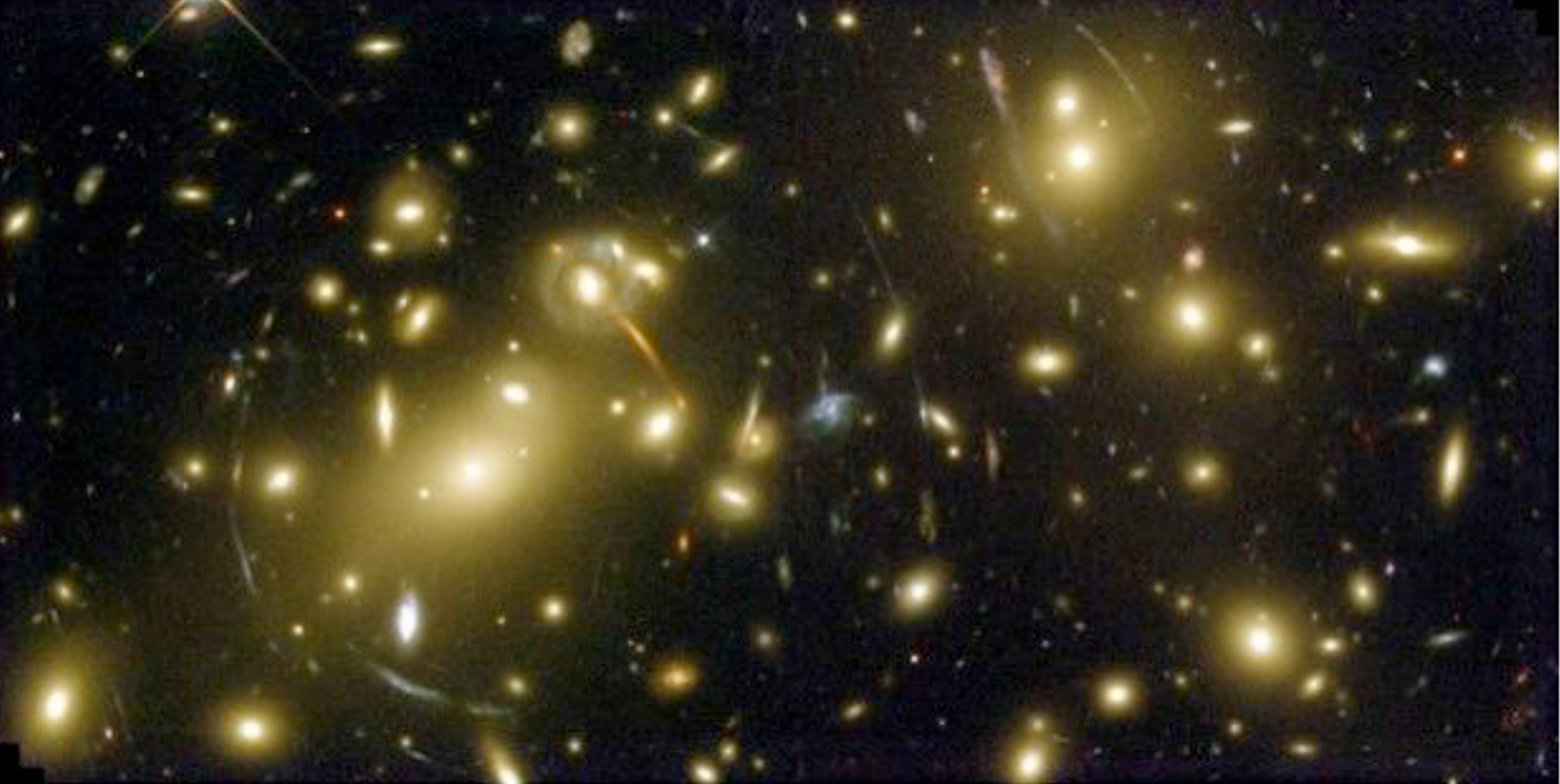}}
\caption{Strong Gravitational Lensing effect observed in the Abell 2218 cluster (W. Couch et al, 1975 - HST).}
\label{abell2218}
\end{figure}

In the beginning of the twentieth century, A. Einstein predicted that massive bodies could be seen as gravitational lenses that bend the path of light rays by creating a local curvature in space-time. One of the first confirmations of Einstein's new theory was the observation during the 1919 eclipse of the deflection of light from distant stars by the sun. Since then, a wide range of lensing phenomena have been detected. The gravitational deflection of light generated by mass concentrations along light paths produces magnification, multiplication, and distortion of images. These lensing effects are illustrated by Fig.~\ref{abell2218} which shows one of the strongest lens observed : Abell 2218, a massive cluster of galaxies some 2 billion light years away towards the constellation Draco. The observed gravitational arcs are actually the magnified and distorted images of galaxies that are about 10 times more distant than the cluster.

\subsection{Gravitational lensing theory}

\begin{figure}[htp!]
\centerline{
\includegraphics[width=9.cm]{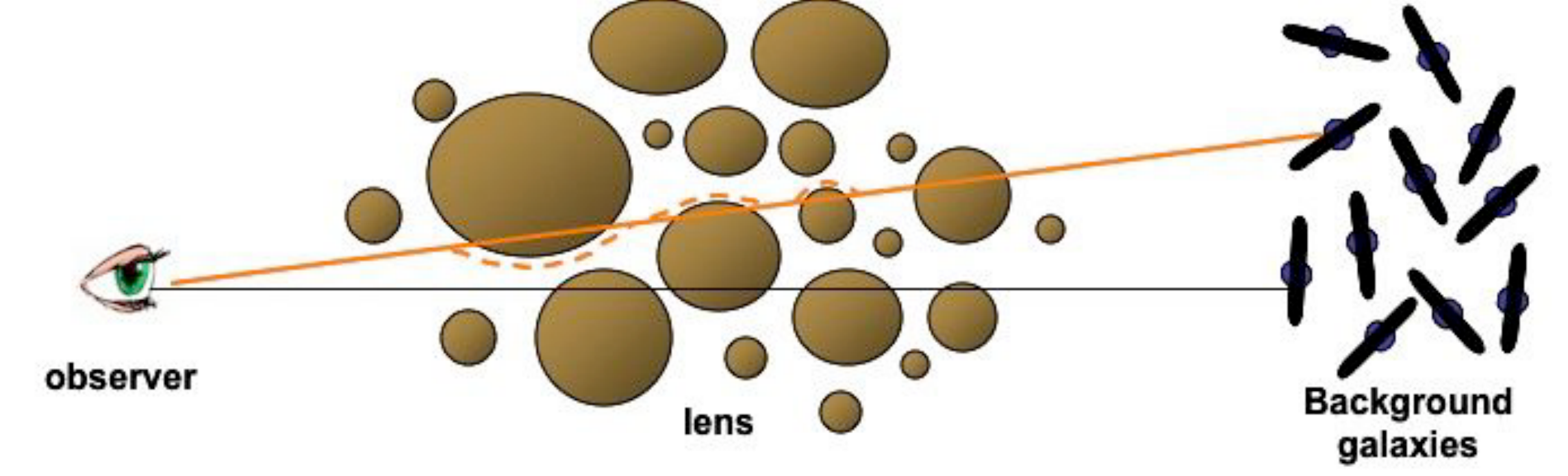}}
\caption{Illustration of the gravitational lensing effect by large scale structures: the light coming from distant galaxies (on the right) traveling toward the observer (on the left) is bent by the structures (in the middle). This bending causes the image of background galaxies to appear slightly distorted. The structures causing the deformations are called gravitational lenses by analogy with classical optics.}
\label{weak1}
\end{figure}

The properties of the gravitational lensing effect depend on all the projected mass density integrated along the line of sight and on the cosmological angular distances between the observer, the lens and the source (see Fig.~\ref{weak1}).\\

%These important observational advances drove theoretical efforts to exploit lensing as an astrophysical tool, establishing gravitational lensing as one of the most dynamic area of research in observational astronomy. 

\subsubsection{The lens equation}

\begin{figure}[htp!]
\centerline{
\includegraphics[width=5.5cm]{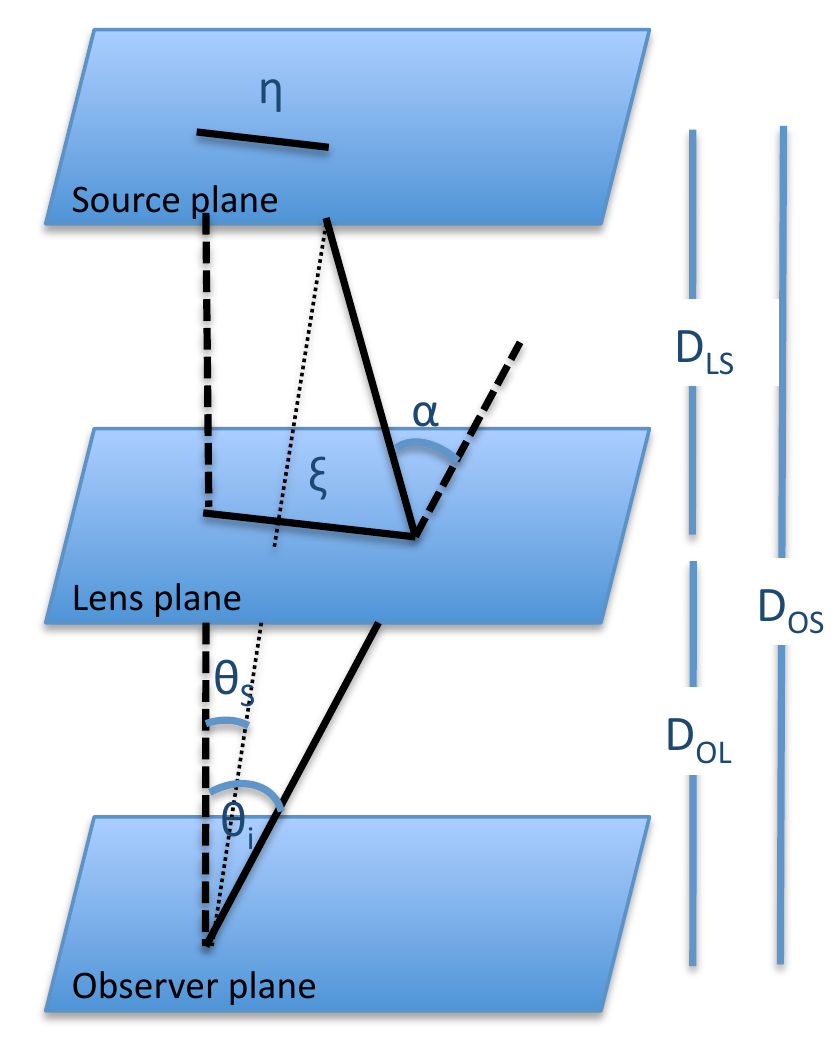}}
\caption{The thin lens approximation.}
\label{im:schema_plan}
\end{figure}

%The light rays that come from the background galaxies are deflected by all the matter integrated along the line of sight. 
In the thin lens approximation, we consider that the lensing effect comes from a single matter inhomogeneity located between the source and the observer. The system is then divided into three planes: the source plane, the lens plane and the observer plane. The light ray is supposed to travel without deflection between these planes with just a slight deflection $\alpha$ while crossing the lens plane (see Fig. \ref{im:schema_plan}).

In the limit of a thin lens, all the physics of the gravitational lensing effect is contained in the lens equation that relates the true position of the source $\theta_S$ to its observed position(s) on the sky $\theta_I$: 

\begin{eqnarray}
\vec{\theta}_S=\vec{\theta}_I-\frac{D_{LS}}{D_{OS}}\vec{\alpha}(\vec{\xi}),
\label{eq:lentille1} 
\end{eqnarray}
with $\vec{\xi} = D_{OL} \vec{\theta_I}$ and 
$D_{OL}$, $D_{LS}$ and $D_{OS}$ are respectively the distance from the observer to the lens, the lens to the source, and the observer to the source. 
The deflexion angle $\alpha$ is related to the projected gravitational potential $\Psi$ obtained by the integration of the 3D Newtonian potential $\Phi(\vec{r})$ along the line of sight:
\begin{eqnarray}
\label{eq:potentielgravi3} 
\vec{\alpha}(\vec{\xi})=\frac{2}{c^2}\int \vec{\nabla}_\bot \Phi(\vec{r}) dz = \vec{\nabla}_\bot \underbrace{\left(\frac{2}{c^2} \int \Phi(\vec{r}) dz,\right)}_\Psi,
\end{eqnarray}
where $c$ is the speed of light and $ \vec{\nabla}_\bot$ is the perpendicular component of the gradient operator.\\

We can distinguish two regimes of gravitational lensing. In most cases, the bending of light is small and the background galaxies are just slightly distorted. This corresponds to the weak lensing effect. Sometimes (as seen previously) the bending of light is so extreme, that the light travels along two different paths to the observer, and multiple images of one single source appear on the sky. For this to happen, the lensing effect must be strong. In this paper, we will only address the weak gravitational lensing regime.\\

\subsubsection{The distortion matrix}
The weak gravitational lensing effect results in both an isotropic dilation (the convergence, $\kappa$) and an anisotropic distortion (the shear, $\gamma$) of the source. To quantify this effect, the lens equation has to be solved. 
Assuming $\theta_I$ is small, a first order Taylor series approximation of the distortion operator, given by the lens equation, can be done :
\begin{eqnarray}
\theta_{S,i} = A_{ij}  \theta_{I, j},
\label{eq:taylor} 
\end{eqnarray}
where $i$ and $j$ correspond respectively to the $i^{th}$ component in the lens plane and the $j^{th}$ component in the source plane and :
\begin{eqnarray}
A_{i,j} = \frac{\partial \theta_{S,i}}{\partial \theta_{I,j}} = \delta_{i,j} -  \frac{\partial \alpha_i (\theta_{I,i})}{\partial \theta_{I,j}} = \delta_{i,j}  -  \frac{\partial ^2 \Psi (\theta_{I,i})}{\partial \theta_{I,i}\partial \theta_{I,j}},
\label{strong2}
\end{eqnarray}
where $A_{i,j}$ are the elements of the matrix $A$ and $\delta_{i,j}$ is the Kronecker delta.
All the first order effects (the convergence $\kappa$ and the shear $\gamma$) can be described by the Jacobian matrix $A$ that is called distortion matrix:
 \begin{eqnarray}
A = (1 - \kappa) \left( \begin{array}{cc}
1 & 0\\
0 & 1\\
\end{array}
\right)
- \gamma \left( \begin{array}{cc}
\cos{2 \varphi} & \sin{2 \varphi}\\
\sin{2 \varphi} & -\cos{2 \varphi}      \\
\end{array}
\right),
\label{strong8}
\end{eqnarray}
where $\gamma_1 = \gamma \cos{2 \varphi}$ and $ \gamma_2 = \gamma \sin{2 \varphi}$. \\

The convergence term $\kappa$ enlarges the background objects by increasing their size, and the shear term $\gamma$ stretches them tangentially around the foreground mass. \\

\subsubsection{The gravitational shear ($\gamma$)}
%The shear $\gamma_1$ and $\gamma_2$ that correspond to the anisotropic distortions can be related to the gravitational potential like this:
The gravitational shear $\gamma$ describes the anisotropic distortions of background galaxy images. 
It corresponds to a two components field $\gamma_1$ and $\gamma_2$ that can be derived from the shape of observed galaxies:
$\gamma_1$ describes the shear in the $x$ and $y$ directions and $\gamma_2$ describes the shear in the $x=y$ and $x=-y$ directions. Using the lens equation, the two shear components $\gamma_1$ and $\gamma_2$ can be related to the gravitational potential $\Psi$ by:
\begin{eqnarray}
\gamma_1 &=& \frac{1}{2}\left[\frac{\partial^2 \Psi(\vec{\theta}_I)}{\partial \theta_{I,1}^2}-\frac{\partial^2 \Psi(\vec{\theta}_I)}{\partial \theta_{I,2}^2}\right] \nonumber\\
\gamma_2 &=&  \frac{\partial ^2 \Psi (\vec{\theta}_{I})}{\partial \theta_{I,1}\partial \theta_{I,2}} .
\label{strong6}
\end{eqnarray}
If a galaxy is initially circular with a diameter equal to 1, the gravitational shear will change this galaxy in an ellipsoid  with a major axis $a = \frac{1}{1-\kappa -|\gamma|}$ and a minor axis $b = \frac{1}{1-\kappa +|\gamma|}$. The eigenvalues of the amplification matrix (corresponding to the inverse of the distortion matrix A) pro- 
vide the elongation and the orientation produced on the images of lensed sources \cite{Mellier 1999}. The shear $\gamma$ is frequently represented by a segment representing the amplitude and the direction of the distortion (see Fig. \ref{kappa0}).\\

\subsubsection{The convergence ($\kappa$)}
The convergence $\kappa$ that corresponds to the isotropic distortion of background galaxy images is related to the trace of the distortion matrix $A$ by:
\begin{eqnarray}
tr(A) &=& \delta_{1,1} + \delta_{2,2} -  \frac{\partial^2 \Psi(\vec{\theta}_I)}{\partial \theta_{I,1}^2}-\frac{\partial^2 \Psi(\vec{\theta}_I)}{\partial \theta_{I,2}^2}, \nonumber \\ 
tr(A) &=& 2- \Delta \Psi(\vec{\theta}_I) = 2(1 - \kappa).
\label{strong4}
\end{eqnarray}

\begin{figure}[htb]
\centerline{
\includegraphics[width=6.5cm, height=6.5cm.]{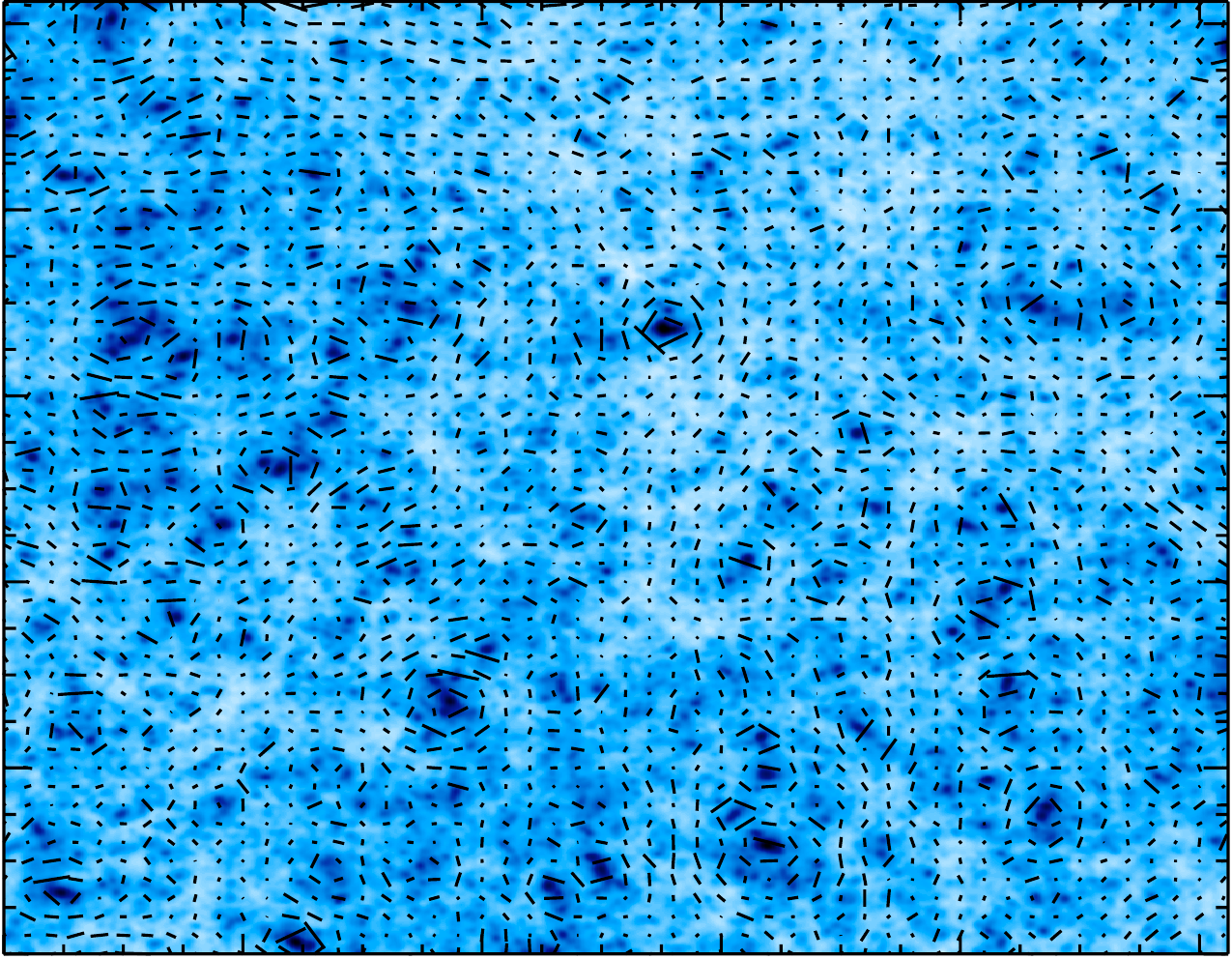}
}
\caption{Simulated convergence map by \cite{Vale 2003} covering a $2^{\circ}$ x $2^{\circ}$ field with $1024$ x $1024$ pixels. The shear map is superimposed to the convergence map. The size and the direction of the segments represent the amplitude and the direction of the deformation locally.}
\label{kappa0}
\end{figure}

The convergence $\kappa$ is defined as half the Laplacian of the projected gravitational potential $\Delta \Psi$ and is then directly proportional to the projected matter density of the lens (see Fig. \ref{kappa0}). For this reason, $\kappa$ is often considered as the mass distribution.\\

\subsection{Weak lensing data analysis}
The dilation and distortion of images of distant galaxies are directly related to the distribution of the (dark) matter and to the geometry and the dynamics of the Universe. As a consequence, weak gravitational lensing offers unique possibilities for probing the statistical properties of dark matter and dark energy in the Universe. In the next sections, we detail the different steps of the weak lensing data analysis along with the different techniques dedicated to these applications. The following sub-problems will be addressed:
\begin{enumerate}
\item Shear estimation from the measurement of the background galaxy ellipticities. 
\item Inversion methods to derive a dark matter mass map from a shear map.
\item Filtering of the dark matter mass map to reduce the noise level.
%\item The mapping of the dark matter distribution
\item Statistical analysis of the weak lensing data to constrain the cosmological model. 
\end{enumerate}
The constraints that can be obtained on cosmology from the weak lensing effect relies strongly on the quality of the techniques used to analyze the data, because the weak lensing signal is very small. In the following, an overview of the different techniques currently used will be given along with future prospects.

\section{Shear estimation}

The weak gravitational lensing effect is so small that it is not possible to detect it from a single galaxy. 
The fundamental problem is that galaxies are not intrinsically circular, so the measured ellipticity is a combination of their intrinsic ellipticity $\epsilon^{int}$ and the gravitational lensing shear $\gamma$. By assuming that the orientation of intrinsic ellipticities of galaxies are random, any systematic alignment arises from gravitational lensing. To estimate the gravitational shear locally, the measurements of many background galaxies must thus be combined.

From this assumption, to perform an estimation of the shear field, we have to correct each galaxy of the field for the Point Spread Function (PSF) due to instrumental and atmospheric effects that distort the apparent shape of background galaxies. A shape determination algorithm has then to be applied to estimate the gravitational shear from background galaxies.

\subsection{Correction of the PSF and shape measurements}
A major challenge for weak lensing is the correction for the PSF. 
Each background galaxy ($S$) is convolved by the PSF of the imagery system ($H$) to produce the 
image that is seen by the instrument ($I^{obs}$):
\begin{eqnarray}
I^{obs}(\theta) = S(\theta)*H(\theta).
\label{psf}
\end{eqnarray}

 This effect can contaminate a true lensing signal and
even for the most modern telescopes, this effect is usually at least of the same order of magnitude 
as the weak gravitational lensing shear, and is often much larger. Then, we must calibrate the PSF using the images of stars.
Indeed, stars present in the field and which correspond to point sources, provide a direct measurement of the PSF, and these can be used to model the variation of the PSF across the field by doing an interpolation between the points where stars appear on the image.
%It is of great importance to characterize accurately the PSF across an image because atmospheric aberration and 
%defects in the detector distort the apparent shape of background galaxies. 
%Correcting for the PSF requires building a model for how it varies across the field. Stars present in the field (that correspond to unresolved objects) provide a direct measurement of the PSF, and these can be used to construct such a model, usually by interpolating between the points where stars appear on the image.
%It is of great importance to characterize accurately the PSF across an image because atmospheric aberration and 
%defects in the detector distort the apparent shape of background galaxies. 

In this section, we briefly review the different methods to correct for instrumental and atmospheric distortions.
These methods are broadly distinguished in two classes. The first class of methods subtract the ellipticity of the PSF from that of each galaxy while the second class of methods attempt to deconvolve each galaxy from the PSF.

The most used method is the KSB+ method that belongs to the first category. This method is the result of a series of successive improvements of the original KSB method proposed by Kaiser, Squires \& Broadhurst \cite{Kaiser 1995}. The core of the method is based on the measurement of the ellipticity of the background galaxies. 
The weighted ellipticity of an object is defined as:
\begin{eqnarray}
\left(\begin{array}{c} \epsilon_1\\ \epsilon_2 \end{array} \right) = \frac{1}{Q_{1,1}+Q_{2,2}} \left(\begin{array}{c} Q_{1,1}-Q_{2,2} \\  2 Q_{1,2} \end{array} \right),
\label{ksb2}
\end{eqnarray}

where:
\begin{eqnarray}
Q_{i,j}=\frac{\int d^2 \theta W(\theta) I(\theta) \theta_i \theta_j}{\int d^2 \theta  W(\theta) I(\theta)}
\label{ksb1}
\end{eqnarray}
are the quadrupole moments weighted by a Gaussian function $W$ of scale length $r$ estimated from the object size, $I$ is the surface brightness of the object and $\theta$ is the angular distance from the object center.
%To provide a gravitational shear estimate $\tilde \gamma$ these galaxy weighted ellipticities parameters have to be corrected for smoothing of %the PSF via the smear susceptibility tensor $P^{sm}$ and calibrated as shears via the shear polarizability tensor $P^{sh}$ both of which %involves higher order shape moments:
%\begin{eqnarray}
%\tilde \gamma = (P^{\gamma})^{-1}[\epsilon -P^{sm}(P^{sm*})^{-1}\epsilon^*],
%\label{ksb3}
%\end{eqnarray}

The PSF correction is obtained by subtracting the star weighted ellipticity $\epsilon_i^*$ from the observed galaxy weighted ellipticity $\epsilon_i^{obs}$.
The corrected galaxy ellipticity $\epsilon_i$ is given by:
\begin{eqnarray}
\epsilon_i = \epsilon_i^{obs} - P^{sm}(P^{sm*})^{-1}\epsilon_i^*,
\label{ksb3}
\end{eqnarray}
where i=1,2 and $P^{sm}$ and $P^{sm*}$) is the smear susceptibility tensors for the galaxy and star, that can be derived from higher-order moments of the images.
%The superscript star indicates that the quantity is computed from the stars.

%An estimate of the local gravitational shear can be obtained by combining the previous ellipticity $\epsilon$ for many nearby galaxies. \\
%where :
%\begin{eqnarray}
%P^{\gamma} = P^{sh} - P^{sm} (P^{sm*})^{-1} P^{sh*}.
%\label{ksb4}
%\end{eqnarray}
%The subscript star signals that the quantity is computed from the stars.\\
This method has been used by many authors although different interpretations of the method have introduced differences between each implementations. One drawback of the KSB+ method is that for non-Gaussian PSFs, the PSF correction is defined poorly mathematically. \\

In \cite{Kaiser 2000}, the authors propose a method to account for realistic PSF better by convolving with an additional kernel to eliminate the anisotropic component of the PSF. \\

The other class of methods attempt a deconvolution of each galaxy from the PSF. The direct deconvolution (of equation \ref{psf}) requires an inversion matrix and that becomes an ill-posed inverse problem if the matrix $H$ is singular (i.e. can not be inverted).
Some methods have been developed to correct for the PSF without a direct deconvolution. %\cite{Rhodes 2000, Massey 2005}. 
These methods try to reproduce the observed galaxies by modeling each unconvolved background galaxy (the background galaxy as it would be seen without a PSF):
\begin{eqnarray}
I^{mod}(\theta) =  S^{mod}(\theta) * H^{*}(\theta).
\label{psfinv2}
\end{eqnarray}

The modeled galaxy ($S^{mod}$) is then convolved by the PSF estimated from the stars  present in the field ($H^{*}$) and the galaxy model is tuned so that the convolved model ($I^{mod}$) reproduces the observed galaxy ($I^{obs}$). One major problem of these methods is that a parametric Gaussian shape is assumed for the PSF and depending on the survey the Gaussian functions are sometimes badly suited to represent PSF shapes. \\

%Recently a direct fully non-parametric deconvolution method has been proposed by \cite{Pichon 2009} based on a maximum a posteriori framework using a smoothness constraint. This method is still at a preliminary stage.

%One major problem of these methods is that a parametric Gaussian shape is assumed for the galaxy model although Gaussian functions are clearly not suited to represent galaxy shapes. New methods, in early stage of development, are attempting to improve the galaxy shape representation. Recently a direct deconvolution method fully non-parametric has been proposed by \cite{Pichon 2009} based on a maximum a posteriori framework using a smoothness constraint. This method is still also at a preliminary stage.

%The study of dark matter reconstructed from weak lensing is strongly constrained by the accuracy with which one can measure galaxy shapes. 
All the methods work by estimating for each galaxy an ellipticity $\epsilon$ (after PSF correction) whose definition can vary between the different methods. 
%An estimate of the local gravitational shear can then be obtained by combining the estimated ellipticity $\epsilon$ for many nearby galaxies. 
%But whatever the method, the measured ellipticity is not a linear function of the applied shear. Then to obtain an unbiased estimator, each method have to determine the relation between their ellipticity and the shear by estimating a shear susceptibility tensor $P^{\gamma}$.
%The shear estimator is then given by the following formula: 
%\begin{eqnarray}
%\tilde \gamma = (P^{\gamma})^{-1} \epsilon
%\label{susceptibility}
%\end{eqnarray}
The accuracy of the shear measurement method depends on the technique used to estimate the ellipticity of the galaxies in presence of observational effects such as noise, poor pixelisation, etc. In the KSB+ method, the ellipticity is derived from quadrupole moments weighted by a Gaussian function. This method has been used by many authors but it is not sufficiently accurate for future surveys. The extension of KSB+ to higher order moments has been done to allow more complex galaxy shapes. Other methods \cite{Bernstein 2002, Kuijken 2006}, based on the ``shapelet" formalism (see \cite{Massey 2005})
% introduced by \cite{Refregier 2003}, 
are more accurate but many shapelet coefficients are needed to represent a galaxy.
%the problem raised in the previous section for PSF representation remains for galaxy representation. 
Indeed, the basis functions of shapelets representation constructed from Hermite polynomials weighted by a Gaussian function are not optimal to represent galaxy shapes that are closer to exponential functions. By consequence, in presence of noise, the accuracy of the shear measurement method based on a shapelet decomposition is not optimal. 
%The best shear measurement method will be based on a sparse representation of the galaxy shape and the PSF. New methods, in early stage of development, are attempting to improve the PSF and galaxie shape representation. 

Many other methods have been developed to address the global problem of the shear estimation. In preparation for the next generation of wide-field survey, a wide range of the shear estimation methods have been compared blindly in the Shear Testing Program (STEP) in order to improve the accuracy of the methods (\cite{Heymans 2006, Massey 2007a}). Several methods have achieved an accuracy of a few percents in the simulated STEP images. However, the accuracy required for future surveys will be greater (of the order of $0.1 \%$) and will require new insights. A new challenge called GREAT08 (\cite{Bridle 2008}) has been recently set outside the weak lensing community as an effort to spur further developments.
%search for new ideas.
%New methods, in early stage of development, are attempting to improve the PSF and galaxy representation in order to use sparsity as a tool to increase the accuracy of the shear measurement methods. 

\subsection{Shear field estimation}
After PSF correction, a catalogue of galaxies can be built with the shape and position of each galaxy present in the field. 
As stated above, the measured shape after PSF correction $\epsilon$  is a combination of the intrinsic ellipticity of the galaxy $\epsilon^{int}$ and the gravitational lensing shear $\gamma$. Thus, the gravitational lensing shear $\gamma$ can only be estimated by averaging over a large number of galaxies.
% in order to have $<\epsilon^{int}> \simeq 0$. 

Usually this shape catalogue is used to characterize the gravitational shear statistically using correlation functions or other statistics (see section \ref{statistic}). It can also be used to map the shear field in order to derive a dark matter mass map (see section \ref{mapping}). 

\begin{figure}[h!]
\centerline{
\includegraphics[width=5.cm,height=5.cm]{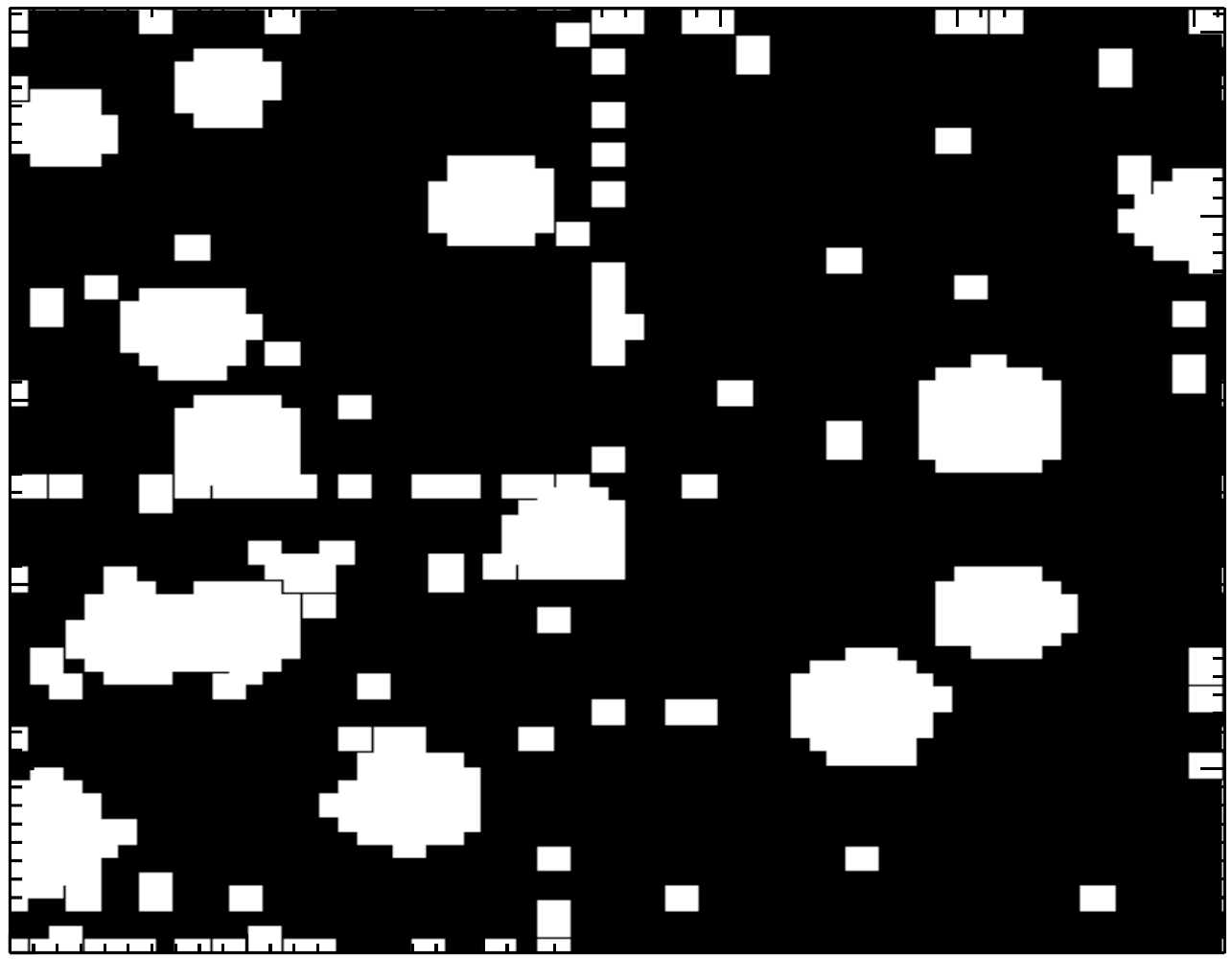}
}
\caption{Mask applied to the CFHTLS data (obtained with the Megacam in the D1 field). The mask covers a  field of $1^\circ$ x $1^\circ$. (J. Berge et al, 2008)}
\label{mask}
\end{figure}

In practice, the shear map is obtained by pixelisation of the field 
%this is done by resampling the field 
in such a way that several background galaxies fall in each pixel. The shear map is then obtained by averaging for each pixel the ellipticity of the galaxies falling into the pixel:
\begin{eqnarray}
\tilde{\gamma}(i,j) = \frac{1}{N(i,j)} \sum_{k=1}^{N(i,j)} \epsilon(x_k,y_k),
\label{shearmap}
\end{eqnarray}
where $\tilde \gamma$ corresponds to the estimated shear for the pixel $(i, j)$ of the shear map, $N(i,j)$ is the number of galaxies within the pixel $(i, j)$ used to estimate the local shear $\tilde \gamma(i,j)$ and $\epsilon(x_k, y_k)$ corresponds to the estimated ellipticity of a galaxy at position $(x_k, y_k)$.
The resulting shear map is subject to some observational effects such as noise that arises both from the measurement error of galaxy ellipticities $\sigma^{meas}$ and the residual of galaxy intrinsic ellipticities $\sigma^{int}$. Another observational effect is the masking out of bright stars from the field that gives a complex geometry to the survey. Fig. \ref{mask} shows an example of mask applied to real data. The analysis of weak lensing data requires to account for these observational effects (see section \ref{sec_inversion} and section \ref{filtering}).

\section{Mapping the dark matter}
\label{mapping}
The problem of mass reconstruction has become a central topic in weak lensing since the 
%early measurements of weak gravitational lensing have led to the 
very first maps of dark matter
%. These maps were the first demonstrations 
have demonstrated that we could see the dark side of the Universe. %without any assumption regarding the light-mass relation, 
%Reconstructing mass maps is also the only way to perform a complete comparison of the dark matter distribution of the Universe as seen in other wavelengths. 

The reconstruction of the dark matter mass map from shear measurements is an ill-posed inverse problem because of observational effects such as noise, complex geometry of the field, etc. This inverse problem can be decomposed in two sub-problems: the reconstruction of the dark matter mass map from the shear field (section IV-A) and the filtering of the dark matter mass map (section IV-B).

\subsection{Weak lensing inversion}
\label{sec_inversion}
\subsubsection{Local inversion}

We first consider the local inversions that have the advantage to address two different problems : the problem of missing data and the finite size of the field.
A relation between the gradient of  $K =\log(1-\kappa)$ and combinations of first derivatives of $g=\frac{\gamma}{1-\kappa}$ have been derived by \cite{Kaiser 1995} :
\begin{eqnarray}
\nabla K  \equiv u \nonumber \\
 \frac{-1}{1-|g|^2}\left( \begin{array}{cc} 1-g_1 & -g_2 \\ -g_2 & 1+g_1 \end{array}\right) \left( \begin{array}{c} g_{1,1} + g_{2,2} \\ g_{2,1} - g_{1,2} \end{array}\right) \equiv u
\label{localinv}
\end{eqnarray}
This equation can be solved by line integration and there exists an infinite number of local inverse formulae which are exact for ideal data on a finite-size field. But they differ in their sensitivity  to observational effects such as noise.
The reason why different schemes yield different results can be seen by noting that the vector field $u$ (the right-hand side of equation \ref{localinv}) has a rotational component due to noise because it comes from observational estimates. 

In \cite{Seitz 1996}, the authors have split the vector field $u$ into a gradient part and a rotational part and they derive the best formula that minimizes the sensitivity to observational effects by convolving the gradient part of the vector field $u$ with a given kernel.

The local inversions reduce the unwanted boundary effects but whatever the formula is used, the reconstructed field is more noisy than that obtained with a global inversion. Another drawback is that the reconstructed dark matter mass map has still a complex geometry that will make the later analysis more difficult. \\

\subsubsection{Global inversion}
A global relation between $\kappa$ and $\gamma$ can be derived, from the relations (\ref{strong6}) and (\ref{strong4}). Indeed, it has been shown by \cite{Kaiser 1993} that the least square estimator $\hat{\tilde \kappa}_n$ of the convergence $\hat{\kappa}$ in the Fourier domain is:
\begin{eqnarray}
\hat{\tilde \kappa}_n(k_1, k_2) =  \hat{P_1}(k_1, k_2) \hat{\gamma}_{1n}(k_1, k_2)+ \nonumber \\
 \hat{P_2}(k_1, k_2)\hat{\gamma}_{2n}(k_1, k_2),
\label{eqn_reckE}
\end{eqnarray}
where the hat symbols denotes Fourier transform and:
\begin{eqnarray}
\hat{P_1}(k_1, k_2)  =  \frac{k_1^2 - k_2^2}{k_1^2+k_2^2}  \textrm{ and } \hat{P_2}(k_1, k_2)  =  \frac{2 k_1 k_2}{k_1^2+k_2^2},
\end{eqnarray}
with $\hat{P_1}(k_1,k_2) \equiv 0$ when $k_1^2 = k_2^2$, and
$\hat{P_2}(k_1,k_2) \equiv 0$ when $k_1 = 0$ or $k_2 = 0$. 
The most important drawback of this method is that it requires a convolution of shears to be performed over the entire sky. As a result, if the observed shear field has a finite size or a complex geometry, then the method can produce artifacts on the reconstructed convergence distribution near the boundaries of the observed field.
%The weak lensing mass inversion problem consists of reconstructing a complete projected mass distribution $\kappa({\mathbf \theta})$ from the incomplete measured shear field $\gamma_i({\mathbf \theta})$ by  using the relation \ref{eqn_reckE}. This is an inverse ill-posed problem that needs to be regularized.
A solution that has been proposed by \cite{Pires 2009} to deal with missing data consists in filling-in judiciously the masked regions by performing an ``inpainting" method simultaneously with a global inversion. Inpainting techniques are an extrapolation of the missing information using some priors on the solution. This new method uses a prior of sparsity in the solution introduced by \cite{Elad 2005}. It assumes that there exists a dictionary $\Phi$ (here the Discrete Cosine Transform) where the complete data are sparse and where the incomplete data are less sparse. The weak lensing inpainting problem consists of recovering a complete convergence map $\kappa$ from the incomplete measured shear field $\gamma_i$. The solution is obtained by minimizing:
\begin{equation}
\min_{\kappa}  \| \Phi^T \kappa  \|_0    \textrm{ subject to }   \sum_i \parallel \gamma_i - M (P_i * \kappa)   \parallel^2 \le \sigma,
\end{equation}
where $\sigma$ stands for the noise standard deviation and $M$ is the binary mask (i.e. $M_i = 1$ if we have information at pixel $i$, $M_i = 0$ otherwise).
This method enables to reconstruct a  complete convergence map $\kappa$ that can be used to do statistic estimation with a good accuracy (see section \ref{statistic}). A comparison with other probes of the matter distribution can also be performed. This comparison is usually done after a filtering of the dark matter map (see section \ref{filtering}) whose quality will be improved by the absence of missing data.

\subsection{Weak lensing filtering}
 \label{filtering}
The convergence map obtained by inversion of the shear field is very noisy even with a global inversion 
%(see Fig. \ref{kappab0}). 
The noise comes from the shear measurement errors and the residual intrinsic ellipticities present in the shear maps that propagate during the weak lensing inversion. An efficient filtering is required to compare the dark matter distribution with other probes.

\subsubsection{Non-Bayesian methods}

%\begin{figure}[htb]
% \centerline{
%\includegraphics[width=5.cm, height=5.cm]{./PDF/shear_noise}}
%\caption{Simulated noisy mass map $\kappa_n$ obtained from a simulated noisy shear map with 100 gal/arcmin$^{2}$ and a shear measurement error  $\sigma^\gamma_\epsilon = 0.3$ arcsec$^{-1}$. }
%\label{kappab0}
%\end{figure}

\begin{itemize}
\item Gaussian filter: \\
The standard method \cite{Kaiser 1993} consists in convolving
the noisy convergence map $\kappa$ with a Gaussian window $G$ with standard
deviation $\sigma_G$:
\begin{equation}
\kappa_G =  G * \kappa_n
                        =   G * P_1 * \gamma_{1n} + G * P_2 * \gamma_{2n}.
\end{equation}
\indent The Gaussian filter is used to suppress the high frequencies of the signal. 
However, a major problem is that the quality of the result depends strongly on the value of $\sigma_G$ that controls the level of smoothing. 
%Fig. \ref{fig_rec_gaus} shows the result obtained with a Gaussian filter from simulated space observations ($\sigma_G = 1$ arcmin).

%\begin{figure}[htb]
%\centerline{
%\includegraphics[width=5.cm,height=5.cm]{./PDF/a_kapint_g8_b2_ops}
%}
%\caption{Mass map reconstruction from simulated weak lensing space observations (Fig. \ref{kappab0}) using a Gaussian filter (with $\sigma=1$ arcmin). The field is  $2^{\circ} \times 2^{\circ}$.}
%\label{fig_rec_gaus}
%\end{figure}

\item Wiener filter:\\
An alternative to Gaussian filter is the Wiener filter (\cite{Bacon 2003, Teyssier 2009}) obtained by assigning the following weight to each $k$-mode:
\begin{equation}
w(k)=\frac{|\hat{\kappa}(k)|^2}{|\hat{\kappa}(k)|^2+|\hat{N}(k)|^2}.
\end{equation}
%\begin{figure}[htb]
%\centerline{
%\includegraphics[width=5.cm,height=5.cm]{./PDF/a_kapint_w_b2_ops}
%}
%\caption{Mass map reconstruction from simulated weak lensing space observations (Fig. \ref{kappab0}) using a Wiener filter. The field is  $2^{\circ} \times 2^{\circ}$.}
%\label{fig_rec_wiener}
%\end{figure}
In theory, if the noise follows a Gaussian distribution, the Wiener filtering provides the minimum variance estimator. However, it is not the best approach, in particular on small scales where non-linear features deviate significantly from gaussianity.
However, Wiener filter leads to reasonable results, generally better than the simple Gaussian filter.\\ 
\end{itemize}

%\subsubsection{Maximum Likelihood estimator}

%The method of maximum likelihood estimates a $\kappa$ that maximizes the likelihood $\mathcal{L}$ (or probability) of obtaining the data $\kappa_n$ given a particular convergence distribution $\kappa$:
%\begin{eqnarray}
%\tilde \kappa=\max_{\kappa} \mathcal{L}(\kappa)
%\end{eqnarray}

%If the noise is uncorrelated and follows a Gaussian distribution, the likelihood term $\mathcal{L}$ can be written:

%\begin{eqnarray}
%\mathcal{L} \propto exp(-\frac{1}{2}\chi^2),
%\label{vraisemblance2}
%\end{eqnarray}
%with :
%\begin{eqnarray}
%\chi^2 = \sum_{x,y} \frac{(\kappa_n(x,y)-\kappa(x,y))^2}{\sigma^2_{\kappa_n}}.
%\label{vraisemblance3}
%\end{eqnarray}

%From a statistical point of view, the method of maximum likelihood is considered to be more robust and yields estimators with good statistical properties. It has been used by \cite{Bartelmann 1996} to reconstruct the gravitational potential. This solution can be improved by adding some prior in the solution.

\subsubsection{Bayesian methods}
\begin{itemize}
\item Bayesian filters \\
Some recent filters are based on the Bayesian theory that considers that some prior information can be used to improve the solution. Bayesian filters search for a solution that maximizes the a posteriori probability using the Bayes' theorem :
\begin{eqnarray}
P(\kappa|\kappa_n)=\frac{P(\kappa_n|\kappa)P(\kappa)}{P(\kappa_n)},
\label{bayes}
\end{eqnarray}
where :
\begin{itemize}
\item $P(\kappa_n|\kappa)$ is the likelihood of obtaining the data $\kappa_n$ given a particular convergence distribution $\kappa$.
\item $P(\kappa_n)$ is the a priori probability of the data $\kappa_n$. This terms, called evidence, is simply a constant that ensures that the a posteriori probability is correctly normalized.
\item $P(\kappa)$ is the a priori probability of the estimated convergence map $\kappa$. This term codifies our expectations about the convergence distribution before acquisition of the data $\kappa_n$.
\item $P(\kappa|\kappa_n)$ is called a posteriori probability. 
\end{itemize}

Searching for a solution that maximizes $P(\kappa|\kappa_n)$ is the same that searching for a solution that minimizes the following quantity $\mathcal(Q)$ :
\begin{eqnarray}
\mathcal{Q} &=&  - \log(P(\kappa|\kappa_n)), \\
\mathcal{Q}&=& - \log(P(\kappa_n|\kappa)) - \log(P(\kappa)).
\label{qteinfo1}
\end{eqnarray}
If the noise is uncorrelated and follows a Gaussian distribution, the likelihood term $P(\kappa_n|\kappa)$ can be written:
\begin{eqnarray}
P(\kappa_n|\kappa) \propto \exp(-\frac{1}{2}\chi^2),
\label{vraisemblance2}
\end{eqnarray}
with :
\begin{eqnarray}
\chi^2 = \sum_{x,y} \frac{(\kappa_n(x,y)-\kappa(x,y))^2}{\sigma^2_{\kappa_n}}.
\label{vraisemblance3}
\end{eqnarray}

The equation (\ref{qteinfo1}) can be expressed as follows:
\begin{eqnarray}
\mathcal{Q} =  \frac{1}{2} \chi^2 - \log(P(\kappa)) = \frac{1}{2} \chi^2 - \beta H,
\label{qteinfo2}
\end{eqnarray}
where $\beta$ is a constant that can be seen as a parameter of regularization and 
$H$ represents the prior that is added to the solution.

If we have no expectations about the convergence distribution, the a priori probability $P(\kappa)$ is uniform and the maximum a posteriori is equivalent to the well-known maximum likelihood. This maximum likelihood method has been used by \cite{Bartelmann 1996, Seljak 1998} to reconstruct the weak lensing field, but the solution needs to be regularized in some way to prevent overfitting the data. 
It has been done via the a priori probability of the convergence distribution.
The choice of  this prior is one of the most critical aspects of the Bayesian analysis.
An Entropic prior is frequently used but there exists many definitions of the Entropy (see \cite{Gull 1989}). One that is currently used is the Maximum Entropy Method (MEM) (see \cite{Bridle 1998, Marshall 2002}). 
%Many priors have been considered such as prior on the Fourier coefficients of the mass distribution (\cite{Squires 1996})
%But it is more natural to assign an Entropy prior (see \cite{Gull 1989}) which is used in the well-known Maximum Entropy Method (MEM). Several authors have applied the MEM method to the weak lensing reconstruction (See \cite{Bridle 1998, Marshall 2002}). 

Some authors (\cite{Bartelmann 1996, Seitz 1998}) have also suggested to reconstruct the gravitational potential $\Psi$ instead of the convergence distribution $\kappa$, still using a Bayesian approach.
But this is clearly better to reconstruct the mass distribution $\kappa$ directly because it allows a more straightforward evaluation of the uncertainties in the reconstruction. \\

%\begin{figure}[htb]
%\centerline{
%\includegraphics[width=5.cm,height=5.cm]{./PDF/a_kapint_mem180btw_b2}
%}
%\caption{Mass map reconstruction from simulated weak lensing space observations (Fig. \ref{kappab0}) obtained with a bayesian filtering: the MEM filtering. The field is  $0.5^{\circ} \times 0.5^{\circ}$ because this code has not been designed for the manipulation of large fields.}
%\label{kappa_MEM}
%\end{figure}

\item Multiscale Bayesian filters \\
A multiscale maximum entropy prior has been proposed by \cite{Marshall 2002} which uses the intrinsic correlation functions (ICF) with varying width. The multichannel MEM-ICF method consists in assuming that  the visible-space image $I$ is formed by a weighted sum of the visible-space image channels $I_j$,  $I = \sum_{j=1}^{N_c} p_j  I_j$ where $N_c$ is the number of channels and $I_j$ is the result of the convolution between a hidden image $h_j$ with a low-pass filter $C_j$, called ICF (Intrinsic Correlation Function) (i.e. $I_j= C_j * o_j$).
In practice, the ICF is a Gaussian. The MEM-ICF constraint is:
\begin{eqnarray}
  H_{ICF} = \sum_{j=1}^{N_c} \mid o_j \mid - m_j
      - \mid o_j \mid \log \left( \frac{ \mid o_j \mid}{m_j}\right).
\end{eqnarray}

Another approach, based on the sparse representation of the data, has been used by \cite{Pantin 1996} that consists in replacing the standard Entropy prior by a wavelet based prior. Sparse representations of signals have received a considerable interest in recent year. The problem solved by the sparse representation is to search for the most compact representation of a signal in terms of linear combination of atoms in an overcomplete dictionary.

The entropy is now defined as :
\begin{equation}
 H(I) = \sum_{j=1}^{J-1} \sum_{k,l} h(w_{j,k,l}).
\end{equation}
In this approach, the information content of an image $I$ is viewed as sum of information at different scales $w_{j}$. The function $h$ defines the amount of information relative to a given wavelet coefficient. Several functions have been proposed for $h$. 

In \cite{Starck 2006}, the most appropriate entropy for the weak lensing reconstruction problem has been found to be the NOISE-MSE entropy
%, defined as :
%\begin{eqnarray}
%h(w_{j,k,l}) =  \int_{0}^{\mid w_{j} \mid } P_n(\mid w_{j,k,l} \mid - u) 
%   (\frac{\partial h(x)}{\partial x})_{x=u} du
%\end{eqnarray}
%where $P_n(w_{j,k,l})$ is the probability that the coefficient
%$w_{j,k,l}$ can be due to the noise: $ P_n(w_{j,k,l}) =
%\mathrm{Prob}(W > \mid w_{j,k,l} \mid) $.
%For Gaussian noise, we have:
%\begin{eqnarray}
% P_n(w_{j,k,l}) & =  & \frac{2}{\sqrt{2 \pi} \sigma_j} 
% \int_{\mid w_{j} \mid}^{+\infty} \exp(-W^2/2\sigma^2_j) dW \nonumber \\ 
% & = & \mbox{erfc}(\frac{\mid w_{j,k,l} \mid }{\sqrt{2}\sigma_j})
%\end{eqnarray}
%and
%\begin{eqnarray}
%h(w_{j,k,l}) = \frac{1}{\sigma_j^2} \int_{0}^{\mid w_{j} \mid} u 
 %             \mbox{ erfc}(\frac{\mid w_{j,k,l} \mid -u}{\sqrt{2} \sigma_j}) du 
%\end{eqnarray}
%The NOISE-MSE penalization 
that presents a quadratic behavior for small coefficients
and is very close to the $l_1$ norm (i.e. absolute value of the wavelet coefficient) when the coefficient
value is large, which is known to produce good results for the analysis of piecewise smooth images. 
The proposed filter called MRLens (Multi-Resolution for weak Lensing) has shown to outperform other techniques (Gaussian, Wiener, MEM, MEM-ICF) in the reconstruction of dark matter. It has been used to reconstruct the largest weak lensing survey ever undertaken with the Hubble Space Telescope. The result is shown Fig. \ref{fig_rec_kappa}. This map is the most precise and detailed dark matter mass map, covering a large enough area to see extended filamentary structures.
%\begin{figure}[htb]
%\centerline{
%\includegraphics[width=5.cm,height=5.cm]{./PDF/a_kapint_fdrl_b2_ops}
%}
%\caption{Mass map reconstruction from simulated weak lensing space observations (Fig. \ref{kappab0}) obtained with a sparse bayesian filtering: the MRLens filtering. The field is  $2^{\circ} \times 2^{\circ}$.}
%\label{fig_rec_kappa}
%\end{figure}
\end{itemize}
\begin{figure}[htb]
\centerline{
\includegraphics[width=6.5cm,height=6.5cm]{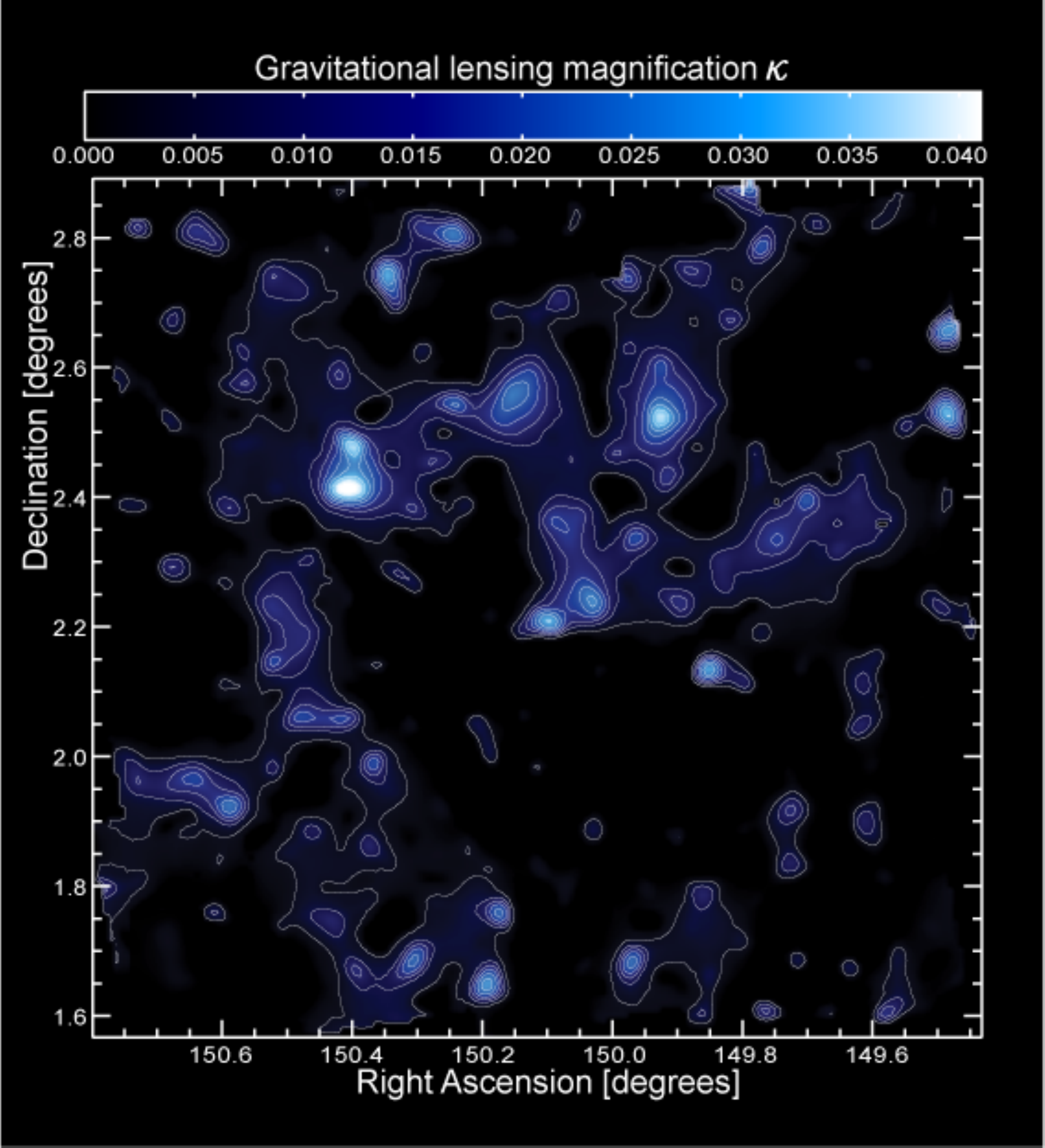}
}
\caption{Map of the dark matter distribution in the 2-square degree COSMOS field by \cite{Massey 2007b}: the linear blue scale shows the convergence field $\kappa$, which is proportional to the projected mass along the line of sight. Contours begin at 0.4 \% and are spaced by 0.5\% in $\kappa$.}
\label{fig_rec_kappa}
\end{figure}

%\end{itemize}

\section{Cosmological model constraints}
\label{statistic}
Image distortion measurements of background galaxies caused by large-scale structures provides a direct way to study the statistical properties of the growth of structures in the Universe. Weak gravitational lensing measures the mass and can thus be directly compared to theoretical models of structure formation. But because, we have only one realization of our Universe, a statistical analysis is required to do the comparison.
The estimation of the cosmological parameters from weak lensing data can be seen as an inverse problem. The direct problem that consists of deriving weak lensing data from cosmological parameters can be solved using numerical simulations. But the inverse problem cannot be solved so easily because the N-body equations used by the numerical simulations can not be inverted. 
A statistical analysis is then required to constrain the cosmological parameters.
The statistical characteristics of the weak lensing field can be quantified using a variety of measures estimated either in the shear field or in the convergence field.
Most lensing studies do the statistical analysis in the shear field to avoid the inversion. But most of the following statistics can also be estimated in the convergence field if the missing data are carefully accounted. 

%Most lensing studies use the two-point statistics of the cosmic shear field because of its potential to constrain the power spectrum of density fluctuations in the late Universe.
%Two-point statistics measure the Gaussian properties of the field. This is a limited amount of information since it is well known that the low redshift Universe is highly non-Gaussian on small scales. Indeed, gravitational clustering is a non linear process and in particular at small scales the mass distribution is highly non-Gaussian. Consequently, using only two-point statistics to set constraints on the cosmological model is limited. An alternative procedure is to consider higher-order statistics of the weak lensing shear field enabling a characterization of the non-Gaussian nature of the signal.

\subsection{Second-order statistics}
\label{twopoint}
%Most lensing studies do the statistical analysis in the shear field to avoid the inversion. But all the following statistics that are currently estimated in the shear field can also be estimated in the convergence field if the missing data are carefully accounted. 
The most common method for constraining cosmological parameters uses second-order statistics of the shear field calculated either in real or Fourier space (or Spherical Harmonic space).
%In general there is an advantage in using Fourier space statistics such as the power spectrum because the modes are independent. 
%The power spectrum $P_{\kappa}(l)$ of the 2D convergence is defined as a function of the modes $l$ by:
%\begin{eqnarray}
%< \hat{\kappa}(\vec{l}) \hat{\kappa}(\vec{l}')> = (2\pi)^2 \delta (\vec{l}-\vec{l}') P_{\kappa}(l),
%\label{cl}
%\end{eqnarray}
%where hat symbols denotes Fourier transforms, $\delta$ is the delta function, the brackets denote an average over $l$.
%$P_{\kappa}(l)$ only depends on $l=|\vec{l}|$ and $\kappa$ is the convergence (i.e. which is proportional to the projected mass distribution).
%\begin{itemize}
%\item The power spectrum $P_{\gamma}$ :\\

The most popular Fourier space second-order statistic is the {\bf power spectrum $P_{\gamma}$} because it can be easily related to the theoretical 3D matter power spectrum $P(k, \chi)$ to estimate cosmological parameters. The correlation properties are more convenient in Fourier space, but for surveys with complicated geometry due to the removal of bright stars, the spatial stationarity is not satisfied and the missing data need proper handling. Consequently, real space statistics are easier to estimate, although statistical error bars are harder to estimate.
%(, but require more computational time). 
%The following two-point statistics can be related to the underlying 3D matter power spectrum via the 2D convergence power spectrum $P_\kappa(l)$:
%\item The shear variance $<\gamma^2>$ :\\
An example of real space second-order statistic is the {\bf shear variance $<\gamma^2>$}, defined as the variance of the average shear $\bar{\gamma}$ evaluated in circular patches of varying radius $\theta_s$. The shear variance $<\gamma^2>$ can be related to the underlying 3D matter power spectrum via the 2D convergence power spectrum $P_{\gamma}$. 
%\begin{eqnarray}
%<\gamma^2> = \int \frac{dl}{2 \pi} l P_{\kappa}(l) \frac{J_1^2(l \theta_s)}{(l \theta_s)^2},
%\label{var}
%\end{eqnarray}
%where $J_n$ is a Bessel function of order $n$.
This shear variance has been used in many weak lensing analysis to constrain cosmological parameters.
%\item The shear two-point correlation function :\\
Another real space statistic is the {\bf shear two-point correlation function $\xi_{i,j}(\theta)$} that is currently used because it is easy to implement and can be estimated even for complex geometry. It is defined as follows :
\begin{eqnarray}
\xi_{i,j}(\theta) = <\gamma_i(\vec{\theta}') \gamma_j(\vec{\theta}' + \vec{\theta})>,
\label{xi}
\end{eqnarray}
where $i, j = 1, 2$ and the averaging is done over pairs of galaxies separated by angle $\theta =|\vec{\theta|}$. By parity $\xi_{1,2} = \xi_{2,1} = 0$ and by isotropy $\xi_{1,1}$ and $\xi_{2,2}$ are functions only of $\theta$.
The shear two-point correlation functions can also be related to the underlying 3D matter power spectrum via the 2D convergence power spectrum $P_{\gamma}$.
%\begin{eqnarray}
%\xi_+(\theta) = \xi_{1,1}(\theta) + \xi_{2,2}(\theta) = \int_0^\infty \frac{dl}{2 \pi} l P_{\kappa}(l) J_0 (l \theta),
%\label{xi2}
%\end{eqnarray}
These two-point correlation functions are the most popular statistical tools used in weak lensing analysis.
%\item The variance of the aperture mass $M_{ap}$ :\\
The {\bf variance of the aperture mass $M_{ap}$} \cite{Schneider 1998} that corresponds to an average shear two-point correlation has been also used in many weak lensing analyses. This statistic is the result of the convolution of the shear two-point correlation with a compensated filter. 
Several forms of filters have been suggested which trade locality in real space with locality in Fourier space.\\
%The variance of the aperture mass can be expressed as a function of the 2D convergence power spectrum by :
%\begin{eqnarray}
%<M_{ap}^2(\theta_s)> = \int \frac{dl}{2 \pi} l P_{\kappa}(l) \frac{576 J_4^2(l \theta_s)}{(l \theta_s)^4},
%\label{xi3}
%\end{eqnarray}
%\end{itemize}

Second-order statistics measure the Gaussian properties of the field. This is a limited amount of information since it is known that the low redshift Universe is highly non-Gaussian on small scales. Indeed, gravitational clustering is a non linear process and in particular at small scales the mass distribution is highly non-Gaussian. Consequently, if only second-order statistics are used to set constraints on the cosmological model, degenerate constraints are obtained between some important cosmological parameters. 
%Fig. \ref{degeneracy} shows the typical constraints that are obtained from second-order statistics on two cosmological parameters $\sigma_8$ and $\Omega_m$. As can be seen, degenerate constraints are obtained between these two parameters. 
 
%\begin{figure}[htb]
%\centerline{
%\includegraphics[width=6.cm]{./PDF/degeneracy}}
%\caption{Cosmological parameter constraints between the density parameter $\Omega_m$ and the amplitude of the matter fluctuations $\sigma_8$. The contours represent the confidence level at $68.3\%$, $95.4\%$ and $99.9\%$.}
%\label{degeneracy}
%\end{figure}

\subsection{Higher-order statistics}
\label{ng}
%In the standard structure formation model, initial random fluctuations are amplified by non-linear gravitational instability to produce a final distribution of mass which is highly non-Gaussian.
%The weak lensing field is highly non-Gaussian: on small scales, we can observe structures like galaxies and clusters of galaxies and on larger scales, we observe some filament structures. 
An alternative procedure is to consider higher-order statistics of the weak lensing shear field enabling a characterization of the non-Gaussian signal. 
%Detecting these non-Gaussian features in weak lensing mass maps can be very useful to constrain the cosmological model parameters. 

The {\bf three-point correlation function $\xi_{i,j,k}$} is the lowest-order statistics which can be used to detect non-Gaussianity.
\begin{eqnarray}
\xi_{i,j,k}(\theta) = <\gamma(\vec{\theta_1}) \gamma(\vec{\theta_2}) \gamma(\vec{\theta_3})>.
\label{xi4}
\end{eqnarray}

In Fourier space it is called {\bf bispectrum} and only depends on distances $|\vec{l_1}|$,  $|\vec{l_2}|$ and $|\vec{l_3}|$:
\begin{eqnarray}
B(|\vec{l_1}|, |\vec{l_2}|, |\vec{l_3}|) &\propto& <\hat{\gamma}(|\vec{l_1}|) \hat{\gamma}(|\vec{l_2}|) \hat{\gamma}^{*}(|\vec{l_3}|)>.
\label{eq:correl3_3}
\end{eqnarray}

It has been shown that tighter constraints can be obtained with the three-point correlation function \cite{Takada 2003}. 

A simpler quantity than the three-point correlation function is provided by measuring the {\bf third-order moment (skewness)} of the convergence $\kappa$ that measures the asymmetry of the distribution. The convergence skewness is primarily due to rare and massive dark matter halos. The distribution will be more or less skewed positively depending on the abundance of rare and massive halos. 
We can also estimate the {\bf fourth-order moment (kurtosis)} of the convergence that measures the peakedness of a distribution.  A high kurtosis distribution has a sharper  ``peak" and flatter  ``tails", while a low kurtosis distribution has a more rounded peak.

\subsection{Other non-Gaussian statistics}
The weak lensing field is highly non-Gaussian: on small scales, we can observe structures like galaxies, groups and clusters and on larger scales, we observe some filament structures. 
Another approach to look for non-Gaussianity is to perform a statistical analysis directly on the non-Gaussian structures present in the convergence field. For example, the galaxy clusters that are the largest virialized cosmological structures in the Universe can provide a unique way to focus on non-Gaussianities present at small scales.
One interesting statistic is the {\bf peak counting} that searches the number of peaks detected on the convergence field corresponding roughly to the cluster abundance.

\subsection{Statistical approach based on sparsity}
It has been proposed by \cite{Pires 2008} to do a statistical analysis based on the sparse representation of the weak lensing data. Several representations have been compared: Fourier, wavelet, ridgelet and curvelet representations. The comparison shows that the wavelet transform is the most sensitive to non-Gaussian cosmological structures. Indeed, by minimizing the number of large coefficient, the wavelet transform makes statistics be more sensitive to the non-Gaussianities present in the weak lensing field. In the same paper, several non-Gaussian statistics have been compared and the peak counting estimated in a wavelet representation, called {\bf Wavelet Peak Counting}, has been found to be the best non-Gaussian statistic to constrain cosmological parameters.

\section{Conclusion}
The weak gravitational lensing effect that is directly sensitive to the gravitational potential provides a unique method to map the dark matter. This can be used to set tighter constraints on cosmological models and to understand better the nature of dark matter and dark energy.  But the constraints derived from this weak lensing effect depend on the techniques used to analyze the weak lensing signal which is very weak. 

The field of weak gravitational lensing has recently seen great success in mapping the distribution of dark matter (Fig. \ref{fig_rec_kappa}). But new methods are now necessary to reach the accuracy required by future wide-field surveys and ongoing efforts are done to improve the current analyses. This paper attempt to give an overview of the techniques of signal processing that are currently used to analysis the weak lensing signal along with future directions. It shows that the weak lensing is a dynamic research area in constant progress.

In this paper, we have detailed the different steps of the weak lensing data analysis thus presenting various aspects of signal processing. For each problem, we have systematically presented a range of methods currently used from earliest to up-to-date methods. This paper shows that a milestone in weak lensing data analysis progress has been the introduction of Bayesian ideas that have provided a way to incorporate prior knowledge in data analysis. The next one could possibly be the introduction of sparsity. Indeed, we have presented new methods based on sparse representations of the data that have already had some success.

\end{document}